\documentclass[showpacs,nobibnotes,prc]{revtex4}
\begin{document}

\title{
Inhomogeneous isospin distribution in the reactions \\
of $^{28}$Si + $^{112}$Sn and
$^{124}$Sn at 30 and 50 MeV/nucleon. 
}

\author{M. Veselsky}
\email{veselsky@comp.tamu.edu}
\thanks{On leave of absence from Institute of Physics of SASc, Bratislava, 
Slovakia}
\author{R. W. Ibbotson}
\thanks{Present address: Brookhaven National Laboratory, Brookhaven, NY 11973}
\author{R. Laforest}
\thanks{Present address: Mallinckrodt Institute of Radiology, 
St. Louis, MO 63110}
\author{E. Ramakrishnan}
\thanks{Present address: Microcal Software Inc, One Roundhouse Plaza, 
Northampton, MA 01060}
\author{D. J. Rowland}
\author{A. Ruangma}
\author{E. M. Winchester}
\author{E. Martin}
\author{S. J. Yennello}
\affiliation{Cyclotron Institute, Texas A\&M University, 
College Station, TX-77843.}

\begin{abstract}
We have created
quasiprojectiles of varying isospin via peripheral reactions of $^{28}$Si + $^{112}$Sn and
$^{124}$Sn at 30 and 50 MeV/nucleon.
The quasiprojectiles have been reconstructed from completely isotopically identified fragments. 
The difference in N/Z of the
reconstructed quasiprojectiles allows the investigation of the disassembly as
a function of the isospin of the fragmenting system. The isobaric yield ratio 
$^{3}$H/$^{3}$He depends strongly on N/Z ratio of quasiprojectiles. 
The dependences of mean fragment multiplicity and mean N/Z ratio of the 
fragments on N/Z ratio of the quasiprojectile are different for light 
charged particles and intermediate mass fragments. 
Observation of a different N/Z ratio of light charged particles and 
intermediate mass fragments is consistent with 
an inhomogeneous distribution of isospin in the fragmenting system. 
\end{abstract}

\pacs{25.70.Mn, 25.70.Pq}

\maketitle

%\section{Introduction}

There has been significant interest in multifragmentation of excited nuclear
matter for many years.  While there has been some success in understanding the
process of multifragmentation and describing this phenomenon in terms of a
liquid gas phase transition \cite{finn}, those efforts have often treated the 
nucleus
as a single component nuclear liquid. In fact, the nucleus is a two component
nuclear liquid.  
Early work by Lamb \cite{lamb} laid the foundations for treating the nucleus 
as a two component system although the results were not connected 
to multifragmentation. Statistical calculations describing multifragment 
disassembly predicted that much of the neutron excess would be observed 
as free neutrons \cite{randrup2}.
Thermodynamic calculations by M\"uller and Serot \cite{MuSe}
lead us to the idea that, for the very neutron rich systems, 
there may exist a distribution of the excited nuclear matter into a neutron 
rich gas and a more symmetric liquid. It is predicted by both lattice gas 
and mean field calculations \cite{chomaz,bali} 
that fragmentation of a system of asymmetric nuclear matter will express 
the characteristics of a two component system. This would result in a second
order phase transition. Also within a dynamical code, constructed to study 
influence of charge asymmetry on spinoidal decomposition of nuclear matter at
sub-saturation densities, differences in the fragmentation are seen as a
function of isospin asymmetry \cite{baran}. 

The difference of the mean N/Z ratio of the light charged particles 
with Z$_f$ $\le$ 2 ( LCPs ) and of the intermediate mass fragments 
with Z$_f$ $\ge$ 3 \hbox{( IMFs )} may be a 
possible experimental signature of a separation into a gas ( resulting mostly 
in emitted LCPs )  and a liquid ( IMFs ). An enhanced production of 
neutron rich H and He isotopes from the neck region has been seen 
experimentally in mid-peripheral collisions \cite{dempsey,sobotka}, while
a favored emission of more symmetric heavy clusters in the mid-rapidity region
has also been shown \cite{ramak}. Quite neutron deficient residues have also 
been seen in intermediate energy reactions \cite{hanold,gonin}.  
Recent results from the reaction of 
$^{112,124}$Sn + $^{112,124}$Sn indicate that the relative abundance 
of free neutrons increases as the
neutron content of the colliding system increases \cite{xu}. The present
work will demonstrate that, in a well defined system with isotopic 
identification of all charged fragments, the mean value of the N/Z ratio 
of LCPs is more sensitive than mean N/Z ratio of IMFs to the N/Z of 
quasiprojectile. The more asymmetric the system is the stronger it will favor 
breaking up into still more asymmetric light fragments 
while the N/Z ratio of heavier fragments remains relatively insensitive. 

%\section{Experiment and Simulation}

This experiment was done with a beam of $^{28}$Si impinging on
$>$1 mg/cm$^{2}$ of $^{112,124}$Sn self supporting
targets. The beam was delivered at energies 30 and 50 MeV/nucleon by the K500
superconducting cyclotron at the Cyclotron Institute of Texas A\&M University.
The detector setup is composed of an arrangement of 68 silicon - CsI(Tl)
telescopes covering polar angles from 1.64$^{o}$ to 33.6$^{o}$ 
in the laboratory frame. Each element is composed of a 300 $\mu$m
surface barrier silicon detector followed by a 3 cm CsI(Tl) crystal. The
detectors are arranged in five concentric rings. The geometrical efficiency is
more than 90\% for each ring. These detectors allow for isotopic
identification of light charged particles and intermediate mass fragments up
to a charge of Z=5. A more detailed description of the detectors and
electronics can be found in ref. \cite{faust}. 
Energy calibration of the silicon detectors was achieved with the use of a
$^{228}$Th source providing 6 calibration points from 5.42 to 
8.78 MeV. The CsI(Tl) detectors were calibrated from the measured energy loss 
in the silicon detector of a given particle of a certain mass and charge and 
from its total kinetic energy calculated with SRIM-96 \cite{srim}.  
In addition, the punch-through points of hydrogen isotopes in CsI(Tl) crystals 
were also used. The overall procedure gives an energy resolution of typically 
3\%. More details of the experimental setup and calibration can be found 
in ref. \cite{laforest}. 

The total charge of quasiprojectile ( QP ) was restricted
to the values Z$_{tot}$=12-15.
We have required that all detected fragments are isotopically
identified. This highly exclusive set of data possesses information on
fragmentation of highly excited projectile-like prefragments over a range in
isospin that can be measured on an event by event basis. 
Simulations have been done with a hybrid code treating the two stages of the
reaction - nucleon exchange and quasiprojectile deexcitation. The first stage
of the calculation was simulated with the Monte Carlo code of Tassan-Got 
\cite{tassangot}. This code implements the model of deep inelastic transfer 
\cite{randrup}.  For the deexcitation of the highly excited quasiprojectile 
we used the statistical model of multifragmentation ( SMM ) \cite{bondorf}. 
The quasiprojectile event sequences generated by the DIT code have been used 
as input of the SMM simulations. The results of the calculation were filtered 
through the FAUST software replica. This hybrid calculation reproduces 
very well the quasiprojectile and fragmentation observables in these 
reactions \cite{sisnprc}. 

%\section{Results and discussion} 

In figure \ref{nznz} we show the effect of including emitted neutrons in the 
calculation of N/Z$_{QP}$. Since the experimental setup cannot detect free
neutrons it is important to see if this might bias the data. 
The N/Z$_{no neut}$ 
is the neutron to proton ratio of the quasiprojectiles simulated using 
hybrid DIT/SMM simulation and filtered using FAUST software replica. 
Thus N/Z$_{no neut}$ applies to quasiprojectiles where only 
charged particles are included. The N/Z$_{neut}$ is the backtraced 
total neutron to 
proton ratio of the primary quasiprojectiles for the same set of events. 
The line represents equal values for N/Z$_{neut}$ and N/Z$_{no neut}$. 
The data presented on figure \ref{nznz} represent the reaction 
$^{28}$Si(50MeV/nucleon)+$^{112}$Sn. While there is some deviation from 
the line due to neutron evaporation, the calculation predicts that N/Z$_{neut}$ 
tracks very well with N/Z$_{no neut}$. The situation is similar 
also when using $^{124}$Sn target or lower projectile energy. 
Therefore for the remainder of this paper N/Z$_{QP}$ will be 
N/Z$_{no neut}$. 

In figure \ref{ib3nz} is shown the dependence of the isobaric yield ratio 
$^{3}$H/$^{3}$He on N/Z$_{QP}$. The yield ratio $^{3}$H/$^{3}$He increases 
as N/Z$_{QP}$ increases. Similar effects have been seen in other systems where 
isotopic or isobaric ratios have been plotted as a function of N/Z of the 
reacting system \cite{dempsey,yennello}. In the present data, the N/Z$_{QP}$ 
is the reconstructed N/Z of all detected charged particles. If one fits 
the data with an exponential function $\exp(a_1+a_2 \hbox{ N/Z}_{QP})$, 
the slope $a_2$ is nearly identical for the reaction with the
$^{112}$Sn target as with the $^{124}$Sn target. The slopes presented 
in table \ref{table1} show that this increase in the yield ratio 
$^{3}$H/$^{3}$He as a 
function of N/Z$_{QP}$ is much larger than would be expected from the change 
in N/Z$_{QP}$ alone. While the integrated values on the two targets are 
different when broken down by the N/Z$_{QP}$ the behavior is similar.  
If one looks at the difference with bombarding energy, the dependence of
yield ratio $^{3}$H/$^{3}$He on N/Z$_{QP}$ is lessened at the higher energy.  
This is consistent with the lattice gas calculation \cite{chomaz} that 
predicted that the $^{3}$H/$^{3}$He ratio would be enhanced at lower
temperatures for the system with excess neutrons.  
At relatively high temperatures above 12 MeV, the ratio would asymptotically
approach the N/Z$_{QP}$ value. The hybrid DIT/SMM calculations reproduce 
the experimental yield ratio $^{3}$H/$^{3}$He rather well. 
This shows that statistical model of multifragmentation 
correctly describes the influence of isospin on the production rates 
of fragments. 
The overall $^{3}$H/$^{3}$He ratio for the reaction
with the $^{112}$Sn target is 0.25 and with the $^{124}$Sn target is 0.35 
at 30 MeV/nucleon
and 0.44 and 0.62 respectively at 50 MeV/nucleon. This increase in the yield
of neutron rich fragments with an increase in the N/Z of the reacting system
is consistent with other studies of isobaric ratios \cite{xu,yennello}. 
What figure \ref{ib3nz} demonstrates is that
this is an integrated feature and merely shows that the system created with
the neutron rich target is more neutron rich. But when one takes the
$^{3}$H/$^{3}$He ratio as a function of the N/Z$_{QP}$, 
the behavior may be seen much more clearly. 

In figure \ref{ib3nzex} we have investigated the effect of excitation energy 
on the $^{3}$H/$^{3}$He ratio. The solid squares represent all events, where 
the excitation energy is less than the mean excitation energy per nucleon 
of the events included in the figure \ref{ib3nz}c ( $^{28}$Si+$^{112}$Sn 
at 50 MeV/nucleon, $\epsilon^{*}$ $\le$ 5.5 MeV/nucleon ). 
The open squares represent the events whose 
excitation energy is greater than the mean excitation energy.  
The slope is steeper for events at lower excitation energies. 
We observed a similar behavior for all reactions regardless of
the bombarding system or beam energy. The results of hybrid calculation 
\hbox{( lines )} are again in reasonable agreement with experiment. 
Further work investigating the excitation energy dependence of 
$^{3}$H/$^{3}$He ratio is forthcoming \cite{mutemp}. 

In figure \ref{mcpnz}a we show the multiplicity of charged particles versus 
N/Z$_{QP }$, again for the reaction $^{28}$Si+$^{112}$Sn at 50 MeV/nucleon.  
The squares represent the multiplicity of all charged particles. 
The multiplicity of charged particles increases as N/Z$_{QP}$ decreases. 
This is then broken down into the multiplicity of light charged particles 
( circles ) and the multiplicity of intermediate mass fragments ( triangles ). 
Here we can see that the multiplicity of LCPs increases with decreasing 
N/Z$_{QP}$. Meanwhile the multiplicity of IMFs increases as N/Z$_{QP}$ 
increases. The situation is similar for all cases. The results are 
identical for different targets at the same projectile energy. The 
dependence of the multiplicity of IMFs on N/Z$_{QP}$ is practically the same 
for all cases. The increase in overall multiplicity of charged particles 
with increasing beam energy is similar to the increase of 
the multiplicity of LCPs.
The results of the hybrid calculation ( lines ) are consistent 
with the experiment. Again the presented information is 
consistent with, but much more illustrative than,
previous works where the difference in isospin of the excited system was only 
approximately known on average \cite{kunde,miller}. Previous works show that 
the multiplicity of IMFs as a function of multiplicity of charged particles 
increases for the more neutron rich system.  While this is true, the present 
data shows that this is concurrent with the decrease in the multiplicity 
of LCPs for neutron rich systems. Our data shows that this effect weakens 
at higher energies, possibly towards the disappearance, that was seen 
by Miller et al. \cite{miller} at much a higher energy. Our data would 
be consistent also with the temperature 
dependence predicted by lattice-gas model calculations \cite{chomaz}. 

In figure \ref{mcpnz}b we show the ratio 
$\frac{N/Z_{IMF}}{N/Z_{LCP}}$ as a function of N/Z$_{QP}$. 
The ratio $\frac{N/Z_{IMF}}{N/Z_{LCP}}$ decreases with increasing 
N/Z$_{QP}$. As there are fewer neutrons available the excess 
protons go into the smaller fragments rather than the larger fragments. 
The least neutron rich quasiprojectiles with N/Z$_{QP}$ $\approx$ 0.5 prefer 
to breakup into very neutron deficient LCPs and much more symmetric IMFs. 
This inhomogeneous isospin distribution washes out rapidly as N/Z$_{QP}$ 
approaches the region of $\beta$-stability. 
Such a behavior may be understood when taking into account that 
the N/Z range of detected LCPs is much wider than the N/Z range 
of IMFs where usually only few stable and nearly stable isotopes are produced. 
The observed inhomogeneous isospin distribution is likely caused 
by more favorable energy balance of the deposition of proton excess into LCPs, 
either free protons or light proton rich clusters 
( see Figs. \ref{ib3nz},\ref{ib3nzex} ). 
When extrapolating this trend towards very neutron rich quasiprojectiles the 
isospin distribution would most likely reappear again in the form of neutron 
rich LCPs and more symmetric IMFs. Such an extrapolation is consistent with 
the predicted asymmetric liquid gas phase transition \cite{MuSe} 
where in addition to the inhomogeneous isospin distribution a spatial 
separation of fractions with different isospin occurs. 
For neutron rich nuclei, such a phase transition is predicted at the values 
of N/Z exceeding 1.5, corresponding to proton concentrations 40 \% 
or less. The data presented here raises the question whether an 
analogous separation of the system into subsystems with 
different isospin may be expected for very neutron deficient 
fragmenting systems. Our data support such a scenario. Nevertheless, one has 
to take into account that the studied fragmenting system is quite small and 
the number of possible exit channels may play an important role in 
the determination of final partition of fragments. 
The influence of the secondary decay was estimated by simulations where 
only an emission of hot fragments took place. The number of simulated 
events, where all fragments have been isotopically resolved, 
decreased for a given overall number of simulated events 
by 50 \% at projectile energy 30 MeV/nucleon and by 90 \% at 50 MeV/nucleon. 
This may be explained by a lower mean multiplicity and 
higher mean mass and charge of hot fragments, what in our case leads 
to a lower number of simulated events where only fragments 
up to boron ( Z$_{f}$ $\le$ 5 ) are emitted. 
Nevertheless, the absence of secondary emission stage did not 
influence the ratio $\frac{N/Z_{IMF}}{N/Z_{LCP}}$ as a function of the 
N/Z$_{QP}$, which remained unchanged within statistical errors. This implies 
that the isospin dependence of the ratio $\frac{N/Z_{IMF}}{N/Z_{LCP}}$ is, 
according to the statistical model of 
multifragmentation, determined in the hot multifragmentation stage.  
The neutron emission does not seem to influence the presented data 
dramatically, as may be seen in figure \ref{nznz}.

%\section{Summary} 
We have demonstrated that systems of varying isospin can be created 
and studied by projectile fragmentation reactions. 
The spread in N/Z of the fragmenting system is larger than the 
difference between the reacting systems.  This can be observed by 
the complete isotopic reconstruction of the quasiprojectile. 
We have investigated the isobaric yield ratio $^{3}$H/$^{3}$He as a function 
of N/Z ratio of the fragmenting system. 
The yield ratio $^{3}$H/$^{3}$He increases 
as N/Z$_{QP}$ increases. This depends more significantly on the fragmenting 
system than on the reacting system. At higher energies this dependence is 
lessened. We conclusively demonstrated that the multiplicity of charged 
particles depends on the isospin of the fragmenting system. As the N/Z$_{QP}$ 
decreases the overall multiplicity of charged particles increases.   
The multiplicity of LCPs increases dramatically while 
the multiplicity of IMFs decreases. This effect is less significant at higher 
bombarding energy. We have also presented evidence of an inhomogeneous 
isospin distribution of the fragmenting system into two fractions with 
different values of N/Z. 

\begin{acknowledgments}

The authors wish to thank the Cyclotron Institute staff for the excellent
beam quality. This work was supported 
in part by the NSF through grant No. PHY-9457376, 
the Robert A. Welch Foundation through grant No. A-1266, and 
the Department of Energy through grant No. DE-FG03-93ER40773.

\end{acknowledgments}

%\section*{References} 

\newpage

\section*{ }

\begin{figure}[!tbp]
\caption{Average N/Z ratio of all fragments emitted from the
quasiprojectile versus those bound in charged fragments for the reaction  
$^{28}$Si(50MeV/nucleon)+$^{112}$Sn. Symbols represent the hybrid DIT/SMM 
calculation. Solid line indicates the case 
\hbox{$N/Z_{neut} = N/Z_{no neut}$.}  
} 
\label{nznz}
\end{figure}

\begin{figure}[!tbp]
\caption{Isobaric yield ratio $^{3}$H/$^{3}$He versus N/Z$_{QP}$ 
for the reactions of $^{28}$Si beam with tin targets.  
(a) - 30MeV/nucleon, $^{112}$Sn, 
(b) - 30MeV/nucleon, $^{124}$Sn, 
(c) - 50MeV/nucleon, $^{112}$Sn, 
(d) - 50MeV/nucleon, $^{124}$Sn.
Symbols represent the experimental data. Lines represent the hybrid 
DIT/SMM calculation.} 
\label{ib3nz}
\end{figure}

\begin{figure}[!tbp]
\caption{
Isobaric yield ratio $^{3}$H/$^{3}$He versus N/Z$_{QP}$ for 
the reaction $^{28}$Si(50MeV/nucleon)+$^{112}$Sn. 
Symbols represent the experimental data 
for two different cuts on apparent excitation energy per nucleon  
( Solid squares - $\epsilon^{*} < \epsilon^{*}_{mean}$,  
open squares - $\epsilon^{*} > \epsilon^{*}_{mean}$ ).  
Lines represent the hybrid DIT/SMM calculation.
( Solid line - $\epsilon^{*} < \epsilon^{*}_{mean}$,  
dashed line - $\epsilon^{*} > \epsilon^{*}_{mean}$ ).  
} 
\label{ib3nzex}
\end{figure}

\begin{figure}[!tbp]
\caption{
(a) - Multiplicity of charged fragments (squares), LCPs (circles) 
and IMFs (triangles) versus 
N/Z$_{QP}$. Corresponding lines represent the hybrid calculation.
(b) - Experimental ratio of the mean values of N/Z of LCPs  
and IMFs versus the N/Z$_{QP}$. 
Data are given for the reaction $^{28}$Si(50MeV/nucleon)+$^{112}$Sn.
} 
\label{mcpnz}
\end{figure}

\begin{table}[!tbp]
%\begin{table}[ht]
\caption{ Values of the slope parameter $a_{2}$ obtained from the fit 
$\log_{10}(\frac{Y(^{3}\hbox{H})}{Y(^{3}\hbox{He})}) 
= a_{1}+a_{2}(\frac{N}{Z})_{QP} $.} 
\label{table1}

\begin{center}

\begin{tabular}{cccccc}
\hline 
\hline 
Reaction & Projectile energy && $a_{2}(exp)$ && $a_{2}(calc)$  \\ 
 & [MeV/nucleon] & & & &\\ \hline
 & & & & & \\
$^{28}$Si+$^{112}$Sn &  30  & & {3.99$\pm$0.39} && {3.89$\pm$0.16} \\  
$^{28}$Si+$^{124}$Sn &  30  & & {4.13$\pm$0.11} && {4.57$\pm$0.24} \\  
$^{28}$Si+$^{112}$Sn &  50  & & {3.16$\pm$0.18} && {3.15$\pm$0.16} \\ 
$^{28}$Si+$^{124}$Sn &  50  & & {2.83$\pm$0.27} && {3.10$\pm$0.11} \\ 
\hline 
\hline
\end{tabular}

\end{center}

\end{table}

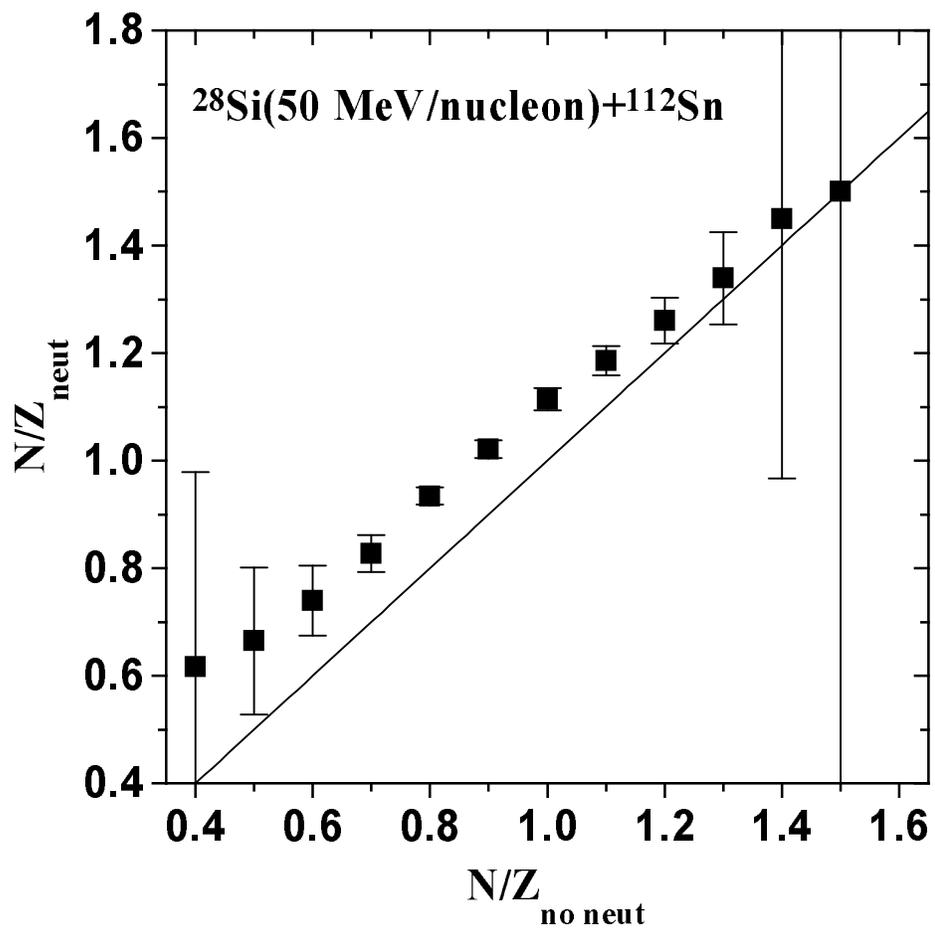
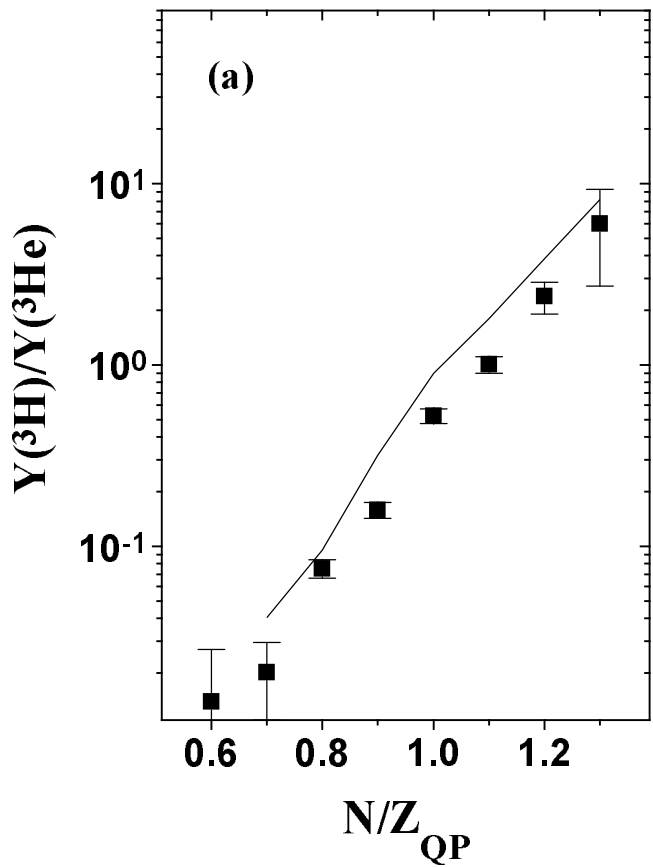
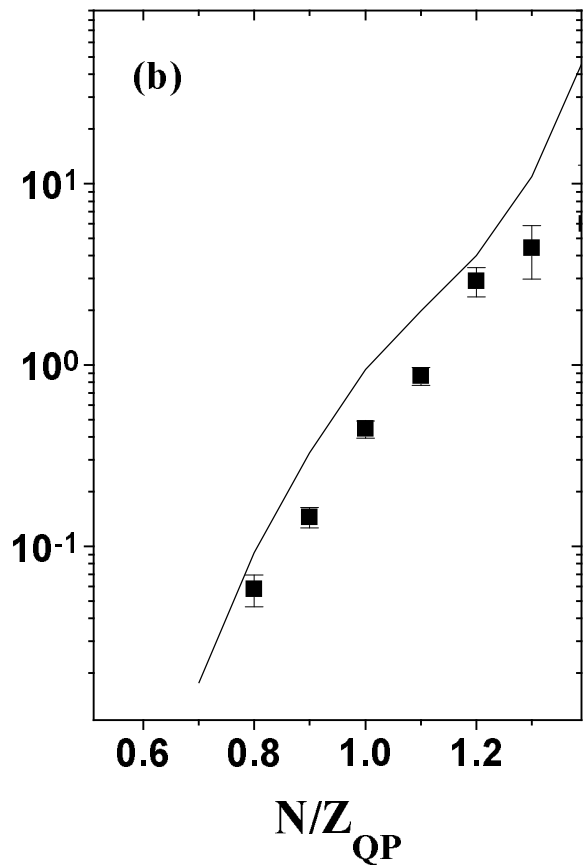
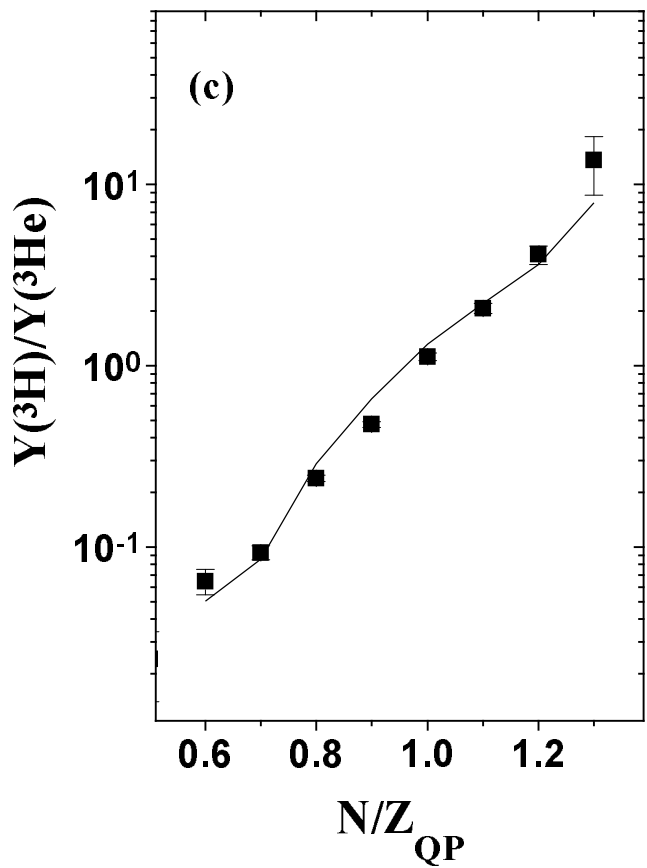
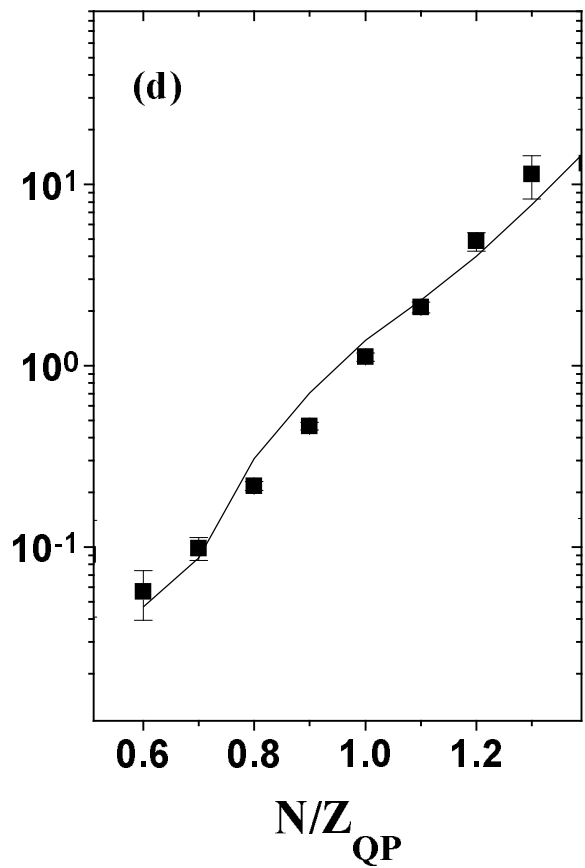
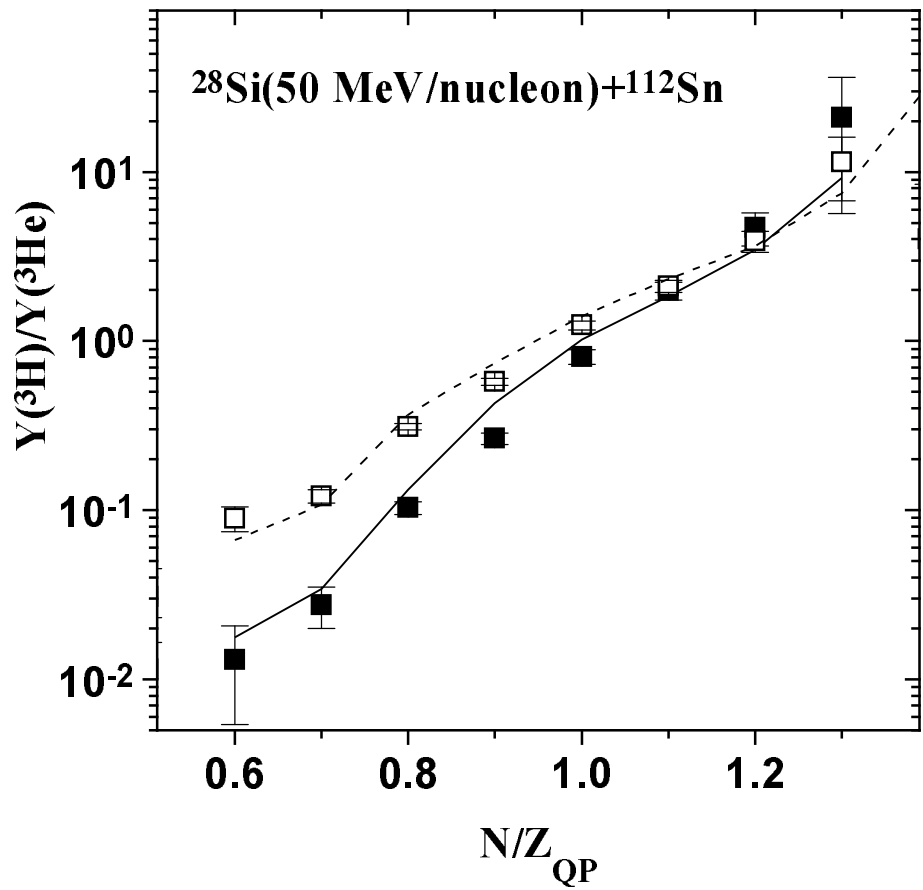
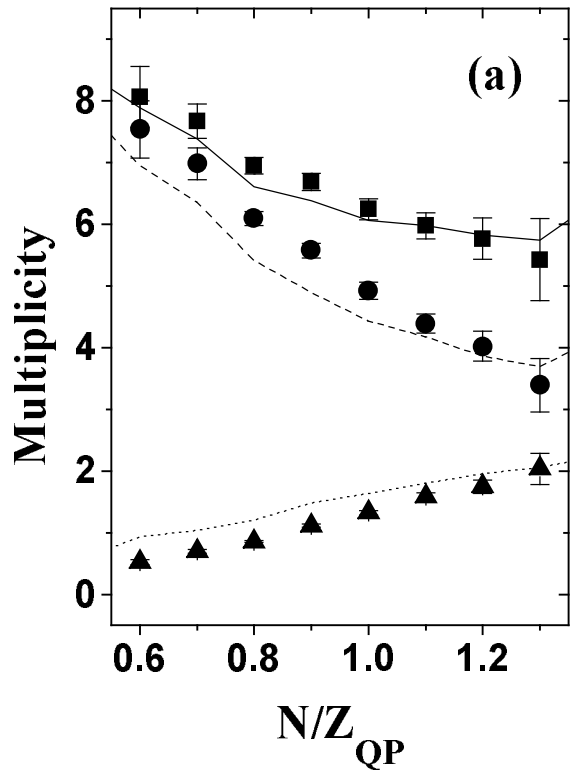
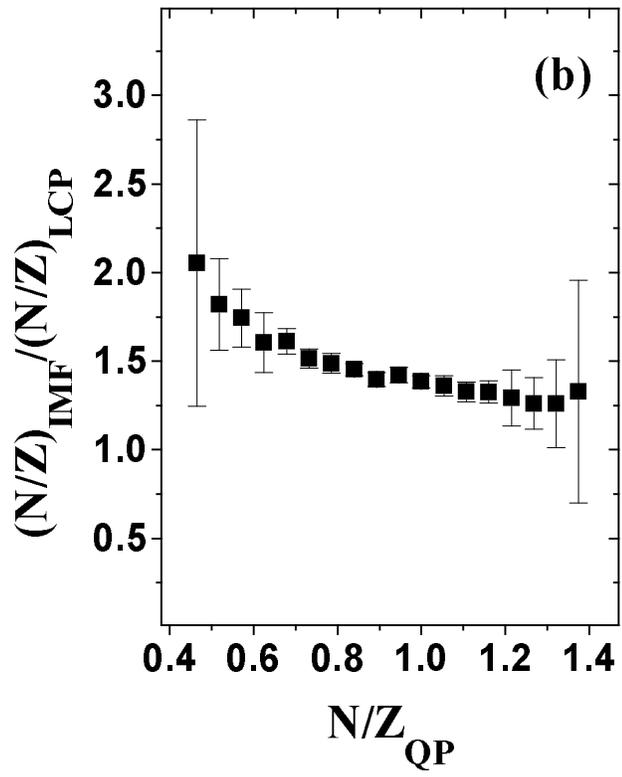
\end{document}